\documentclass[aps,twocolumn,preprintnumbers,amsmath,amssymb]{revtex4}
\usepackage{graphicx}
\usepackage{epsfig}
\usepackage{amsmath}
\usepackage{graphicx}
\usepackage{dcolumn}
\usepackage{bm}
\usepackage{amssymb}
\usepackage{amsmath}
\usepackage{epsf}
\usepackage{subfigure}
\usepackage{epstopdf}

\usepackage{pifont}
\usepackage{color}
\usepackage{wasysym}

\begin{document}

\newcommand{\pr}{\partial}
\newcommand{\rta}{\rightarrow}
\newcommand{\lta}{\leftarrow}
\newcommand{\ep}{\epsilon}
\newcommand{\ve}{\varepsilon}
\newcommand{\p}{\prime}
\newcommand{\om}{\omega}
\newcommand{\ra}{\rangle}
\newcommand{\la}{\langle}
\newcommand{\td}{\tilde}

\newcommand{\mo}{\mathcal{O}}
\newcommand{\ml}{\mathcal{L}}
\newcommand{\mathp}{\mathcal{P}}
\newcommand{\mq}{\mathcal{Q}}
\newcommand{\ms}{\mathcal{S}}

\newcommand{\nl}{$\newline$}
\newcommand{\nll}{$\newline\newline$}

\newcommand{\Fa}{F\left[\begin{array}{c|c} x & x^\p \\ t+\Delta t & t \end{array} \right]}
\newcommand{\Fb}{F\left[\begin{array}{c|c} x_t & x_0 \\ t & t_0 \end{array} \right]}
\newcommand{\Fby}{F\left[\begin{array}{c|c} y_t & y_0 \\ t & t_0 \end{array} \right]}

\newcommand{\vspa}{\vspace{2mm}}
\newcommand{\vspb}{\vspace{3mm}}
\newcommand{\vspc}{\vspace{4mm}}

\title{How and why does statistical mechanics work}

\author{Navinder Singh\footnote{criticism/comments are welcomed}}
\email{navinder.phy@gmail.com}
\affiliation{Physical Research Laboratory, Navrangpura, Ahmedabad-380009, India.}

\begin{abstract}
As the title says we want to answer the question; how and why does statistical mechanics work?  As we know from the most used prescription of Gibbs we calculate the phase space averages of dynamical quantities and we find that these phase averages agree very well with experiments. Clearly actual experiments are not done on a hypothetical ensemble they are done on the actual system in the laboratory and these experiments take a finite amount of time. Thus it is usually argued that actual measurements are time averages and they are equal to phase averages due to ergodicity.  Aim of the present review is to show that ergodicity is not relevant for equilibrium statistical mechanics (with Tolman and Landau). We will see that the solution of the problem is in the very peculiar nature of the macroscopic observables and with the very large number of the degrees of freedom involved in macroscopic systems as first pointed out by Khinchin. Similar arguments are used by Landau based upon the approximate property of "Statistical Independence". We review these ideas in detail and in some cases present a critique. We review the role of chaos (classical and quantum) where it is important and where it is not important. We criticise the ideas of E. T. Jaynes who says that the ergodic problem is conceptual one and is related to the very concept of ensemble itself which is a by-product of frequency theory of probability, and the ergodic problem becomes irrelevant when the probabilities of various micro-states are interpreted with Laplace-Bernoulli theory of Probability (Bayesian viewpoint).

In the end we critically review various quantum approaches (quantum-statistical typicality approaches) to the foundations of statistical mechanics. The literature on quantum-statistical typicality is organized under four notions (1) kinematical canonical typicality, (2) dynamical canonical typicality, (3) kinematical normal typicality, and (4) dynamical normal typicality.
Analogies are seen in  the Khinchin's classical approach and  in the modern quantum-statistical typicality approaches. 
\end{abstract}

\maketitle


"Unless the conceptual problems of a field have been clearly resolved, you cannot say which mathematical problems are the relevant ones worth working on, and your efforts are more than likely to be wasted"

---- E. T. Jaynes.

\tableofcontents

\section{Introduction}

We start the introduction by considering the following views of great men:
\nll
"Although nobody is in doubt today of the validity of the remarkable interpretation of thermodynamics with which statistical mechanics, following the efforts of Boltzmann and Gibbs, has recently provided us, it still remains extremely difficult to give a completely accurate justification for it"

----Louis De Broglie\cite{Broglie}.

\nll
"The problem [defining the ensemble distribution depending upon the external conditions imposed on a system] was solved by Gibbs, although a rigorous justification of the distributions obtained is a complicated problem that is still not completely solved at the present time. It is not even clear to what extent this rigorous justification is possible"

----D. N. Zubarev\cite{Zubarev}.
\nll
"There is probably no other well-established field of theoretical physical science that is as much plagued by paradox and criticism of its fundamental logic as is statistical mechanics"

----Joseph Edward Mayer and Maria Goeppert Mayer\cite{JEMayer}

\nll

"If we are describing only a state of knowledge about a single system, then clearly there can be nothing physically real about frequencies in the ensemble; and it makes no sense to ask, "which ensemble is the correct one?"... Gibbs understood this clearly; and that, I suggest, is the reason why he does not say a word about ergodic theorems,..."

						--- E. T. Jaynes\cite{ETJaynes1}

\nll

There is lot of confusion regarding ergodic hypothesis in the literature (regarding whether it is necessary for statistical mechanics or not) and in the usual discussions of people. Also the reasons how and why statistical mechanics works are not properly understood.  The problem appears complicated as one has to take into consideration various ingredients of vary different nature (from the character of experimental measurements, large number of degrees of freedom, and the microscopic dynamical properties) involved in the statistical approach for dynamical systems. In the present review we try to dis-entangle various ingredients and present the resolution of the ergodic problem in a simpler and clear way.

A detailed  study of literature shows that there are mainly three camps at the foundations of statistical mechanics (1) ergodic school, (2) non-ergodic school,  and (3) quantum foundations of statistical mechanics.  We critically analyse all the three approaches to reach on a broader understanding of the foundations of statistical mechanics or we attempt to see the "complete picture".

Aim of the present review is to show that ergodicity is not relevant for equilibrium statistical mechanics of macroscopic systems (with Tolman and Landau). We will see that the solution of the problem is in the very peculiar nature of the macroscopic observables (they are sum functions) and with the very large number of the degrees of freedom involved as first pointed out by Khinchin. Similar arguments are used by Landau based upon the approximate property of "Statistical Independence". We critically review these ideas. We also review the role of chaos (classical and quantum) in the foundations of statistical mechanics, i.e., where it is important and where it is not important. For students and beginners the purpose of the article is best served if they first understand the first chapter of Landau-Lifshitz's book on statistical physics and read the little book of Khinchin on the Mathematical foundations of statistical mechanics. {\it Again our motivation is to have a broader view of the subject and a pedagogical presentation. }

In the present section we present a brief summary of the essential topics in the form of questions and answers.  In section II we review the ergodic approach to the foundations of statistical mechanics by dividing it into further subsections. First defining the ergodic problem and then giving a historical account of how the ergodic problem came up with the works of  Maxwell and Boltzmann. In subsection II(c) the reformulation of the ergodic problem by Birkhoff is given and the resolution of the ergodic problem by Khinchin is given in subsection D. The role of chaos, integrability, and non-integrability is discussed in subsection E.

In section III we present the non-ergodic approach. This section consists of two subsections one on Landau-Lifshitz's approach and another on E. T. Jaynes approach with a critique. We also advance the plausibility arguments for equal-a-priori probability hypothesis.

The last section is devoted to the quantum foundations of statistical mechanics. In this we review the eigenstate thermalization hypothesis, the quantum ergodic theory of von Neumann, and other recent quantum-statistical typicality approaches. We will see that von Neumann's quantum ergodic theorem is a general statement applicable to systems with many degrees of freedom  and the eigenstate thermalization hypothesis is a consequence of quantum chaos, other approaches falling under typicality properties are also analyzed.

The issue of compatibility of microscopic time reversibility and macroscopic time irreversibility is not considered here. This has been clarified in detail in the beautiful papers of Jeol Lebowitz\cite{lebowitz}(for irreversibility in quasi-isolated systems see\cite{yukalov}).

\subsection{Is ergodic hypothesis necessary for the foundations of statistical mechanics?}

In their famous book\cite{LandauLifshitz} Lev Landau wrote:
"In the discussion of the foundations of statistical physics, we consider from the start the distribution of small subsystems......this allows the complete avoidance of Ergodic or similar hypotheses, which are in fact not essential as regard to these aims [i.e., the foundations of physical statistics]"
\nll

As we know from the most used prescription of Gibbs we calculate the phase space averages of dynamical quantities using appropriate ensembles and we find that these phase averages agree very well with experiments. Clearly actual experiments are not done on a hypothetical ensemble they are done on the actual system in the laboratory and these experiments take a finite amount of time. Thus it is usually argued that actual measurements are time averages and they are equal to phase averages due to ergodicity,

\begin{equation}
\bar{f} =\lim_{T\rightarrow \infty} \frac{1}{T}\int_0^T f(x(t))dt = \int \rho(x) f(x) dx.
\end{equation}

Now we will give the following reasons  against ergodicity.

(1) If one assumes that during the measurement time the system samples all the microstates and $T\rightarrow \infty$ limit of equation (1) is reached during the measurement time,  then ones assumption is wrong. It is a well know fact that time taken by the system to sample all the available phase is fantastically large (more than the age of the universe). For example\cite{jaynesL4}, if we consider a nanoparticle with $1000$ nuclei each with spin half and consider only the nuclear spin system. The total number of quantum state are $2^{1000}$. Consider that these spins continuously flip from up spin to down spin and vice versa and this spin flipping is caused by phonons from the system and bath in which the nanoparicle is present. At room temperature this frequency is about $10^{12}$ cycles per second. Let us assume that each spin makes a transition during this period. Thus the total number of transitions per second is $1000\times 10^{12}= 10^{15}$. The time taken by the system to go over all the states is $\frac{2^{1000}}{10^{15}}\sim 10^{286} ~~sec$, and the age of universe is $10^{17} ~sec$. {\it Thus clearly during the laboratory measurement time the full ensemble is not realized, and only a very tiny fraction of the total number of the members of the ensemble is realized.} Note that the time scale diverges nearly exponentially with the number of atoms in the sample.

(2) If one accepts ergodic hypothesis (and do not accept explanation (1)), that the laboratory measurements are time averages (measurement time scale being much large as compared to microscopic dynamical time scale (time taken to flip a spin $\sim 10^{-12}~sec$), then one accepts equation (1) and accepts $T\rightarrow \infty$), then time average value of an observable (or a result of measurement) in a series of experiments (repeating the same experiment again and again) on the same system will agree with ensemble average. This is in fundamental contradiction with what we observe i.e. thermodynamical measurements are not strictly reproducible\cite{Tolman}. Fluctuations in thermodynamic measurements are the unavoidable consequence of our lack of control of microscopic dynamics and thus our lack of control of the initial micro-condition at the start of the measurement.

\subsection{How does statistical mechanics work and why does statistical mechanics work so well?}

The above discussion against ergodicity opens the question "but how does statistical mechanics work?" and the theory has been so successful. The answer is:

In statistical mechanics we concern with macroscopic systems and with macroscopic observables which are very special ones. Special in the sense that they are "sum functions" i.e., their value for the whole system is equal to the sum of equivalent functions for the small parts of the body (for example, energy of the whole system is equal to the sum of energies of the small parts of the body). Due to the  property called "statistical independence" of small parts of the body, the sum functions take almost constant value on the energy hypersurface (details are given in section II(D)). 
\nll
The microscopic time taken by the measurement produces that "same" value. Thus equality of time average and phase space average is imposed due to the fact of "temporal constancy of special macroscopic observables in equilibrium". The magnitude of fluctuations goes as $\la O^2 \ra - \la O \ra^2 \propto 1/\sqrt{N}$. Thus these are very small for macroscopic systems.

Put differently, for an extremely large number micro-states (called typical micro-states) the macro-state of the system is the same. Only very exceptional micro-states ("bad states" with fantastically small probability) give non-typical behaviour. To explain these lines let us consider a macroscopic system and measure an observable. Repeat the experiment many many times, each time the starting micro-state is different but important point is that each time we get approximately the same value for the observable (since huge number of micro-states are equilibrium micro-states). It is extremely rare that we get a value during a measurement which is very different from the value in other measurements.

This "typicality" is at the heart of why statistical mechanics explain thermodynamic behaviour. Here the probability theory inters into statistical description. The typicality explains its high predictability for macroscopic systems (even though statistical mechanics has probabilistic basis).

\subsection{What is the role of microscopic dynamics in equilibrium statistical mechanics ?}

As the macro-observables are extremely insensitive to micro-condition and microscopic dynamics, the role of microscopic dynamics is very insignificant in equilibrium statistical mechanics.  See section II. D. for a detailed explanation, and the role of Fermi-Pasta-Ulam problem in this connection (section II. E).

Also it is good idea to abandon 19th century philosophy of explaining everything we see around with microscopic dynamics called "reductionism". Boltzmann initially tried to explain 2nd law of thermodynamics with microscopic molecular dynamics but later on realized the need of of statistical laws. At each level of complexity new laws emerge not explainable "purely" by Shroedinger equation\cite{anderson}.

\section{Ergodic approach}

\subsection{The Ergodic Problem}

In standard practice, for example for a system of volume $V$ with $N$ particles in equilibrium with very large heat bath at temperature $T$, one use's Gibbs canonical ensemble

\begin{equation}
\rho_{can}(p,q) = \frac{e^{-\beta H(p,q)}}{Z(\beta, V,N)},~~ Z(\beta,V,N) = \int e^{-\beta H(p,q)} dpdq.
\end{equation}
   
Here $\rho_{can}(p,q) dpdq$ is the probability to find the system in the infinitesimal phase volume $dpdq$ around the phase point $(p,q)$ [ $(p,q) \equiv (\bf{p}_1,p_2,...,p_N ; q_1,q_2,...,q_N)$ for a system of $3N$ degrees of freedom]. $\beta = \frac{1}{k_B T}$ with $k_B$ called Boltzmann constant.

The negative logarithm of partition function $Z$ is proportional to the free energy of the system

\begin{equation}
F(\beta,V,N) = - \frac{1}{\beta} ln Z(\beta,V,N).
\end{equation}

From the free energy and its derivatives all the needed information is extracted. Each time we get the Gibbs ensemble averaging, for example, pressure is given as

\begin{equation}
P = - (\frac{\pr F }{\pr V})_{\beta,N} = -\frac{1}{Z}\int e^{- \beta H} (\frac{\pr H }{\pr V})_{\beta,N} dpdq = <P>.
\end{equation}

With Hamiltonian $H$ as a function of the volume $V$ (through potential function with boundaries etc.).

Thus our computational algorithm involves phase space averaging. But the actual experiments are done on the given system in the laboratory (not on the hypothetical ensemble).  Measurements during the experimentation take finite amount of time and thus what we measure in laboratory is the time averages not the ensemble averages. So the immediate question arises how to justify the replacement of time averages with ensemble average. This is called the ergodic problem.

There is a considerable confusion regarding the precise meaning of ergodic hypothesis.
Thus we will present a brief historical account about the origin and its mis-interpretations by Ehrenfests\cite{ehrenfests,brush}. In this analysis we will better understand the ergodic program for justifying statistical mechanics.

\subsection{Historical account: Boltzmann's ergodic program}

Boltzmann's first attempt to reduce the second law of thermodynamics to a theorem in mechanics appeared in 1866 and his last papers in columns of Nature in 1890s regarding his debates on irreversibility. During his scientific career  (1866-1895) Boltzmann tried to understand the kinetic theory of gases and thermodynamics with  many different approaches based on atomic constitution of matter. Favouring one approach at one time and then rejecting it for the another and then returning back again to the previous one. This is a very distinctive character of Boltzmann\cite{cer}.  Roughly, from 1866-1871 he was interested in deriving second law of thermodynamics purely from mechanics (in his 1866 paper no probabilistic argument is present), in 1867 he reads Maxwell's work and his subsequent papers from 1868 to 1871 uses probabilistic notions after Maxwell's velocity distribution function. In these papers he extends Maxwell's results to a gas in an external potential (Maxwell-Boltzmann law). In the paper\cite{boltzmann68} of 1868 he first introduces his ergodic hypothesis. He was considering an isolated gas in an arbitrary initial state ( see Ehrenfests'\cite{ehrenfests}), then he argues--based on the empirical fact that systems tend to equilibrium and permanently stay there--that average behaviour over a long time interval will the thermal equilibrium behaviour (note that this long before his H-theorem). According to Ehrenfests here he introduces the concept of ensembles (much before Gibbs). In justifying the equivalence of ensemble averages and temporal averages he introduces a hypothesis (note that he do not use the word ergodic at this point, word Ergoden appears much later in his 1884 paper in a different context). Thus the present day terminology is largely due to Ehrenfests. Note also that this is the birth of modern statistical mechanics. The older treatments in which statistical aspects are related to single molecules are called (by Ehrenfests) the "Kineto-statistics of the molecule" and the 1868 paper treats gas model as a whole and Ehrenfests call it "Kineto-statistics of the gas model".  

Important point to be noted here is that the version of ergodic hypothesis which is attributed to Boltzmann is stronger than what the founding father has thought: 

"The great irregularity of the thermal motion, and the multiplicity of forces that act on the the body from outside, make it probable that the atoms themselves, by virtue of motion that we call heat, pass through all possible positions and velocities consistent with the equation of kinetic energy, and that we can therefore apply the equations previously developed to the coordinates and velocities of the atoms of warm bodies"

\begin{figure}
\includegraphics[height = 3.5cm, width = 5cm]{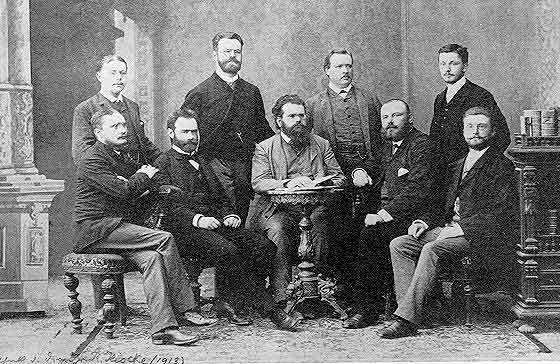}
\caption{Boltzmann in the center.}
\end{figure}

Now the present strong form (trajectory passing through each and every point of the energy surface for an isolated system) of ergodic hypothesis cannot be attributed to Boltzmann, it is the mis-interpretation on the Ehrenfests side. The representative phase point of the system passing through every point of the phase space is impossible. In fact later in 1913 the impossibility of ergodic systems (stronger form) was proved by independently by Rosenthal and Pancherel on measure-theoretic arguments. Ehrenfests after mis-interpreting the Boltzmann ergodic hypothesis proposed some what weaker form and they called it the quasi-ergodic hypothesis,  which meant that the trajectory of phase point covers the energy surface densely even though not actually passing through every point of it. An excellent discussion of this mis-interpretation is given in the work of S. G. Brush\cite{brush}.

In the period 1872-1878 he wrote two most important papers of his life, the 1872 paper contains what we now call "Boltzmann equation" and H-theorem. Here he claimed that his H-theorem provided the desired theorem from mechanics to explain irreversibility and second law of thermodynamics. This came under a serious objection due to Loschmidt in 1876 who insisted on the incompatibility of the time-asymmetric behaviour (shown by H-theorem)  and the time-symmetric behaviour of microscopic equations of motion (These debates are now well understood and documented\cite{lebowitz}). As a result Boltzmann rethought the basis of his approach and presented a very different approach in 1877 what we now call permutational argument and his famous result $S \propto ln W$ (see appendix A) written in modern form by Max Planck. Equilibrium is now conceived as most probable macro-state instead of stationary macro-state (in time). It is highly probable for a system initially in a non-equilibrium state to move towards equilibrium state (real space density distribution in accord with external conditions(potentials) and Maxwellian velocity distribution) as the phase space volume of the equilibrium macro-state is fantastically large as compared to non-equilibrium state. Thus the equilibrium is most probable state but arbitrary deviations from it are also probable, but it turns out that the probability is fantastically small. This originates the notions of typicality.

Third period, in 1880s he left the "purely" probabilistic approach and again went back to his mechanical explanation of the laws thermodynamics (we see here that Boltzmann did not stick to one approach). He wrote an important paper in 1884 largely forgotten by now (see the review by Gallavotti\cite{boltz84}).  This paper is a generalization of a paper by Helmholtz on the mechanical models of thermodynamics based on monocyclic systems. Boltzmann proves a theorem called heat theorem, i.e., $\frac{dU +PdV}{T} ~is~~exact$. Here $T$ is the time average of the kinetic energy $K$, $U= K+\Phi$ is the total energy, and $\Phi$ the potential energy which depend on an experimentally controllable parameter $V$. The motivation here was to find mechanical analogues of thermodynamic entropy. Result was proved for monocyclic systems by Helmholtz and Boltzmann generalize this for high dimensional systems under the assumption of ergodicity (note that this is the second time he introduces ergodic considerations). In his discrete world-view he imagined the phase trajectory visiting all the  discrete cells of the phase space. With the introduction of some stationary phase  space distribution (he calls it "monode"), the calculation of difficult time averages can be replaced with much simpler phase averages.  He considers an ensemble of systems in exactly same macroscopic conditions and a family of stationary distributions which he calls "monode" (note that this the introduction of ensembles much before Gibbs). A monode which verifies the heat theorem is called by him the "orthode". Here (first time) he considers two orthodes (1) ergode ($\equiv $ microcanonical ensemble) and (2) holode ($\equiv$ canonical ensemble), see fig 1.

\begin{figure}
\includegraphics[height = 5cm, width = 5cm]{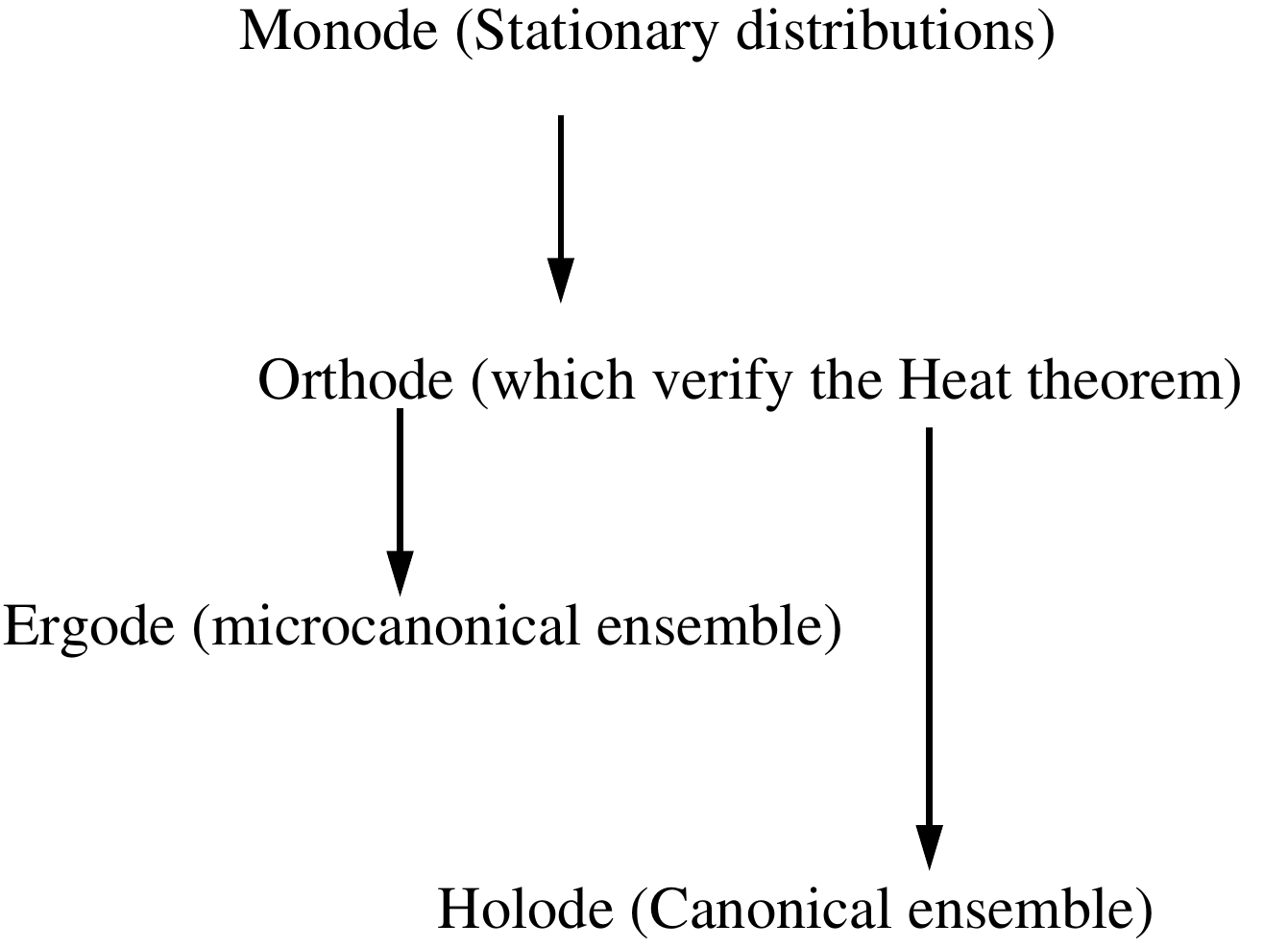}
\caption{Boltzmann's 1884 formulation of statistical ensembles.}
\end{figure}

As we have seen in section I(A), in real thermodynamic measurements we do not have infinite time averages, and thus we are not justified in equating the finite time averages with infinite time averages and then using the ergodic hypothesis (which has never been proved in generality). The answer to this was already clear to Boltzmann. The answer lies in the peculiarities of the thermodynamic observables and the large number of degrees of freedom involved. We will discuss this in section II(D).

We also note that the original ergodic hypothesis as envisaged by Boltzmann was a weaker statement as compared to modern version of that.


\subsection{Ehrenfests, Birkhoff and the modern ergodic theory}

Ehrenfests' 1911 encyclopedia article\cite{ehrenfests} generated lot of interest about ergodic systems in the community of mathematicians (Paul Ehrenfest was the student of Boltzmann).
\nll
If one is able to prove the ergodic hypothesis for a given system i.e., $\bar{f} =\lim_{T\rightarrow \infty} \frac{1}{T}\int_0^T f(x(t))dt = \int \rho(x) f(x) dx.$, then
\begin{dinglist}{42}
\item one eliminates the necessity of determining initial state of the system and of solving Hamilton's equations.

\item one justifies the dynamical foundation of statistical mechanics. \large{[\smiley]}

\end{dinglist}

Clearly, then statistical mechanics reduces to a branch of mechanics. But story is not so simple. It is a very difficult problem. In 1931, G. D. Brikhoff\cite{bri} reduced the ergodic property of a dynamical system to an equivalent property called "metric transitivity". He proves the following theorems:

\begin{dinglist}{45}

\item $\bar{f} = \lim_{T\rightarrow \infty} \frac{1}{T}\int_0^T f(U^t x(0))dt$ exists for almost every $x(0)$.

\item A necessary and sufficient condition for the system to be ergodic is that the phase space be metrically transitive.
\end{dinglist}

A system is "metrically in-transitive" iff there exists regions $X_1$ and $X_2$ of phase space such that $X_1 \cap X_2 = \emptyset$ and $X_1 \cup X_2 = X$, which are invariant under the system's dynamics: $U^t X_1 \subseteq X_1 $ and $U^t X_2 \subseteq X_2 $ for all $t$.

In simple words phase point wanders all the available phase space if and only if the system is metrically transitive.
\nll
This property again cannot be experimentally verifiable. Thus the implications of Brikhoff's theorems for physical statistical mechanics are inconclusive. 
\nll
But luckily  or unluckily,  we have seen that ergodicity is not required for doing statistical mechanics of macroscopic systems.

\subsection{Resolution by Khinchin}

\begin{figure}
\includegraphics[height = 4cm, width = 4cm]{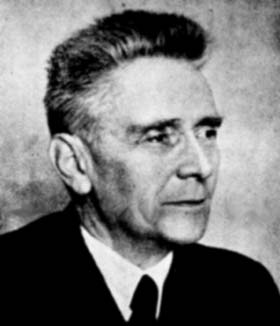}
\end{figure}

Khinchin\cite{khinchin}(see also the 4th chapter of the book\cite{bookvul}) approached the ergodic problem from physical point of view. He pointed out the importance of the following points 

\begin{enumerate}

\item in statistical mechanical systems of interest the number of the degrees of freedom is very large.

\item in statistical mechanical systems the observables of interest are very special ones they are "sum functions" ($f(p,q) = \sum_{i = 1}^N f_i(p_i,q_i) $).

\end{enumerate}

With these considerations he proves the following two theorems:
\nll
Theorem (1): Consider a physical quantity $f_i$ pertaining to a single molecule and define phase coefficient of correlation: $R(s) = \frac{\overline{f_i(p_i,q_i;t) f_i(p_i,q_i;t+s)}}{\overline{f_i^2(p_i,q_i)}}$.  If $R(s)\rta 0$ for $s \rta \infty$ the function $f_i(p_i,q_i)$ is ergodic ($\hat{f_i}(p_i,q_i)~(time~average) = \bar{f_i}~(phase~average)$).

\vspb
The physical assumption that goes into this theorem is:
"Because of the fact that the given system consists of a very large number of molecules it is natural to expect that knowledge of the state of a single molecule  at a certain moment does not permit us to predict anything (or almost anything) about the state in which this molecule will be found after a sufficiently long time"
-----Khinchin.
\nl
This is equivalent to the idea of molecular chaos. 
\vspb

Next he proves a much more relevant result (to physical statistics):\nl
Theorem (2)\cite{bookvul}: 
\begin{equation}
\boxed{Prob\left(\frac{|f-\bar{f}|}{|\bar{f}|} \geq k_1 N^{-1/4}\right) \leq k_2 N^{-1/4}}
\end{equation}
Here $f$ is the actual value of the observable (no time averaging !) pertaining to the whole system, and $k_1$ and $k_2$ are $O(1)$. He proves that the correlation co-efficient between phase functions is actually very small for large $N$.
\nll
This implies that the physically relevant observable are self averaging!

\vspb
The above statement is the resolution of the whole problem {\it(for macroscopic systems and sum-function observables).}

\subsection{Role of chaos, integrability, and non-integrability}

As far as one is concerned with macroscopic systems (large $N$) and sum function observables chaos, integrability, and non-integrability does not play any role in the foundations of statistical mechanics. As usual statistical mechanical systems in the laboratory are never isolated, the molecular chaos assumption of Khinchin seems to be true and the consequences are clear to us.

\vspb
But for "isolated" low dimensional systems and non-sum-function observables the case is not so simple. Before 1955 there was a general consensus that weak extraneous interactions makes the statistical mechanical systems ergodic (phase point wandering the whole of phase space).
 
\vspb
In 1955, Fermi and his collaborators did a numerical experiment on a chain of non-linearly coupled harmonic oscillators. Their expectation was that the small non-linear coupling will make the system ergodic (energy stored initially in one of the normal modes will go over in time to all other modes). But the result was quite surprising. Consider the Hamiltonain of the system as  $H = \sum_{i=0}^N \left(\frac{p_i^2}{2 m} +\frac{K}{2} (q_{i+1}-q_i)^2 + \frac{\ep}{r} (q_{i+1}-q_i)^r \right)$. For $\ep=0$ system is completely integrable and the energy $E_k = \frac{1}{2} (\dot{Q}_k^2 + \omega^2_k Q^2_k)$ of each normal mode $Q_k = \sqrt{\frac{2}{N}} \sum_{n=1}^N q_n(t)sin(\pi k n/N)$ is conserved.

It was expected that when $\ep\neq 0$ energy stored initially in one normal mode will "spread" to all other modes. But this did not happen, the energy periodically comes back to the original mode showing no sign of equipartition of energy.

\vspb
This "paradox" can be explained with KAM theorem. Which states that if $\ep$ is small enough then on the constant energy surface, invariant tori survive in a region whose measure tends to $1$ as $\ep\rta 0$.

\vspb
There is a threshold $\ep_c$ called KAM threshold, if (1) $\ep<\ep_c$ KAM tori play a major role and system does not follow equipartition, and (2) if $\ep>\ep_c$ then KAM tori has a minor role and equipartition happens.

\subsubsection{Canonical ensemble in FPU system}

In 1987, Livi, Pettini, Ruffo, and Vulpiani published an important paper\cite{livi} on the relevance of chaos regarding the predictions of canonical ensemble.
\begin{figure}
\includegraphics[height = 1cm, width = 8cm]{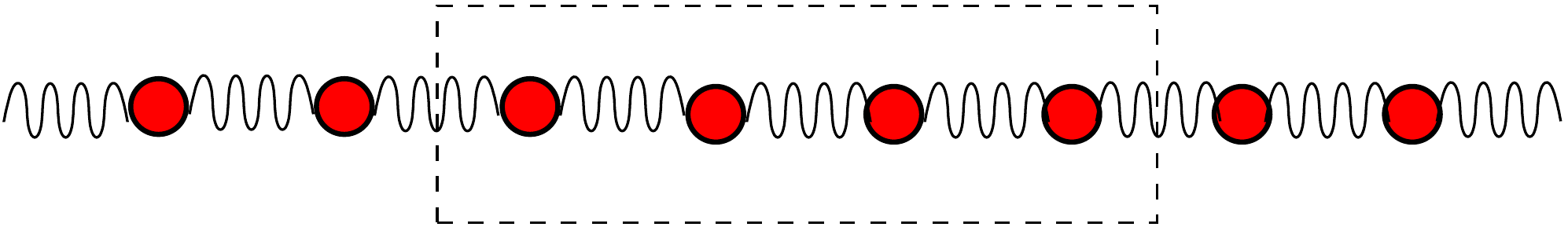}
\caption{Canonical ensemble in FPU system}
\end{figure}
They divided the chain into coupled subsystems containing large number of particles (see figure 3). Let $U$ is the per particle time averaged energy of the subsystem $= \hat{E(t)}/N_{sub}$ and $C_v$ it's heat capacity $\frac{\hat{E(t)^2} - \hat{E(t)}^2}{N_{sub} T^2},~~~T = \hat{p^2}$.

\vspace{3mm}
It was found that the time averages\footnote{The time required for averaging in the KAM tori regime is very large and diverges as one reach $\ep\rta 0$ limit (private communication with Roberto Livi)} agrees well with the predictions of the canonical ensemble, even though the system was having the KAM tori. However the story is not so simple. They also considered another model of coupled rotators in which only $U$ showed canonical behaviour not $C_v$ as the integrability-to- non-integrability parameter was changed (see for details chapter 4 of \cite{bookvul}). Note also that $C_v$ is not a sum function.

Thus one can loosely say that {\it "more coarse your observables, less you depend on chaos".}

\section{Non-Ergodic approach}

\subsection{Landau-Lifshitz approach}

As expressed at the beginning of section I (A), Lev Landau did not believe in the need of ergodic or similar hypotheses in justifying the foundations of statistical mechanics\cite{LandauLifshitz}. From the beginning they consider a system always present in a bath, i.e., subsystem in a given system which is in itself present in bigger system (as in nature no system is ideally isolated).  They argue that the extreme complexity of the interactions of the system with the surrounding bodies makes the system's phase point to wander in the phase space. Let in a sufficiently long interval of time $T$ the phase point spends $\Delta t$ amount of time in a given volume $\Delta p \Delta q$ of the phase space, then,

\begin{equation}
w = \lim_{T \rta \infty}\frac{\Delta t}{T}
\end{equation}  

is the probability that, if the system is observed at any arbitrary time, its phase point will be found in the volume $\Delta p \Delta q$. Going to the infinitesimals $dw = \rho_{actual}(p,q) dpdq$. Here $\rho_{actual}(p,q)$ is the density of that temporal probability distribution, i.e., $\rho_{actual}(p,q) dpdq$ is the probability to find the system, observed at any time, in the infinitesimal phase volume $dpdq$.

The important point to be noted here is that we have "only one" system under consideration (no statistical ensemble is considered at this point). The distribution defined $\rho_{actual}(p,q)$ is a temporal distribution for that system.  

With the introduction of statistical distribution function one can calculate the mean value of any dynamical quantity $f(p,q)$ as
\begin{equation}
\bar{f} = \int f(p,q)\rho_{actual}(p,q) dpdq.
\end{equation}
It is obvious by the definition (6) of probability that the statistical averaging (equation (7)) is exactly equivalent to the infinite time averaging

\begin{equation}
\bar{f} = \lim_{T\rta\infty}\frac{1}{T}\int_0^T f(t)dt.
\end{equation}  

Thus avoiding any ergodic hypothesis, as there is no ad-hoc introduction of any probability distribution.  The $\rho_{actual}(p,q)$ is the actual probability distribution in the system's phase space from its temporal behaviour.

But it is non-trivial to prove that 

\begin{equation}
\lim_{T\rta\infty}\frac{1}{T}\int_0^T f(t)dt = \int f(p,q)\rho_{microcanonical}(p,q) dpdq.
\end{equation}

This has never been proved in generality (it turns out that this is a very difficult problem for a system with large number of degrees of freedom). We will see from Landau's argument and with the assumption of "statistical independence" that $\rho_{actual} \simeq \rho_{microcanonical}$ when number of degrees of freedom involved becomes very large.

Then they argue\cite{LandauLifshitz} that predictions of statistical mechanics are very reliable due to the fact that the relevant observables take almost constant value on the energy surface. They also consider "sum" functions $f = \sum_i^N f_i)$. Their considerations are based on a very general fact of "statistical independence" of various parts (pertain to subsystems) of the macroscopic observables (it is not true for long range interacting systems). On "statistical independence" they prove: $\frac{\sqrt{\la \Delta f^2\ra}}{\la f\ra} \propto \frac{1}{\sqrt{N}}$. One can say
\nll
$\left( \begin{array}{ccc}
Statistical~~independence & \equiv & molecular~~chaos\\

 of~~Landau-Lifshitz   &  &  of~~Khinchin

\end{array} \right)$

\subsubsection{Landau's argument: shortest derivation of the canonical ensemble}

\begin{enumerate}

\item The distribution function of two sub-systems is equal to the product of individual sub-systems functions (statistical independence) , i.e., $\rho_{12} = \rho_1\rho_2$ (for simplicity of notation here $\rho_{actual} \equiv \rho$). 
\item But $log (\rho_{12}) = log(\rho_1) + log(\rho_2)$. Thus log of the distribution function is an additive integral-of-motion.

\item Due to Liouville's theorem (Hamiltonian dynamics) distribution function is an integral-of-motion (temporal invariance of the phase space distribution of an ensemble).

\item We have only seven independent additive integrals of motion $E(p,q), {\mathbf{P}}(p,q)$, and ${\mathbf{M}}(p,q)$(from mechanics).

\item $\Longrightarrow log(\rho) = \alpha + \beta E(p,q) + {\mathbf{\gamma . P}(p,q) + {\mathbf{\delta . M}}(p,q)  }$

\item From which canonical ensemble ensues $\rho = e^{\alpha + \beta E(p,q)}$, say, for a system in a box.

\end{enumerate}

The important point to be noted is that there are $6N-7$ other integrals of motion (excluding the additive ones, i.e., energy, three components of linear and three components of angular momentum). These remaining integrals of motion are not additive thus they do not play any role in the Landau's argument.

Thus we see in Lev Landau's program for the foundations of statistical mechanics, if one starts from the beginning with a system present in a bigger system and one assumes the property of statistical independence and take note of the additive integrals of motion then one can reach the canonical distribution and no ergodic hypothesis is required as we have not constructed any ensemble.

\subsubsection{Why the hypothesis of equal-a-priori probabilities is plausible?}
We observe that the property of statistical independence is equivalent to the hypothesis of equal-a-priori probabilities. We give two reasons:

Reason (1): There are two views at the  foundations of statistical mechanics (1) is based on the principle of equal-a-priori probabilities (Kubo's book, Tolman's etc) and the (2) is based on quasi-closed subsystems and the additive integrals of motion (due to L. D. Landau, see their book). If one analyze them carefully one sees that both view-points are consistent with each other, first one comes from the second.

\vspb
The principle of equal-a-priori probability is just a consequence of the fact that there are mechanical invariants of motion proportional to log of the distribution function for a quasi-closed subsystem  $(E \propto \log \rho)$ (due to statistical independence), which is another integral of motion due to Liouville's theorem, thus one has  $\log(\rho) =  \alpha + \beta E_{subsystem}$.  The right hand side of this equation is the content of dynamics (conservation laws)\footnote{say for a system in a rigid fixed box.}. The left hand side consists of a more subtle quantity related to statistical properties and a consequence of the theorem of dynamics and the property of statistical independence. Since $E(p,q)$ is constant no matter where is phase point is in the available phase space. Consequently log of $\rho$, thus $\rho$ is same for all possible $(p,q)$. This is an equal-a-priori probabilities (see figure 4).
\begin{figure}
\includegraphics[height = 2cm, width = 5cm]{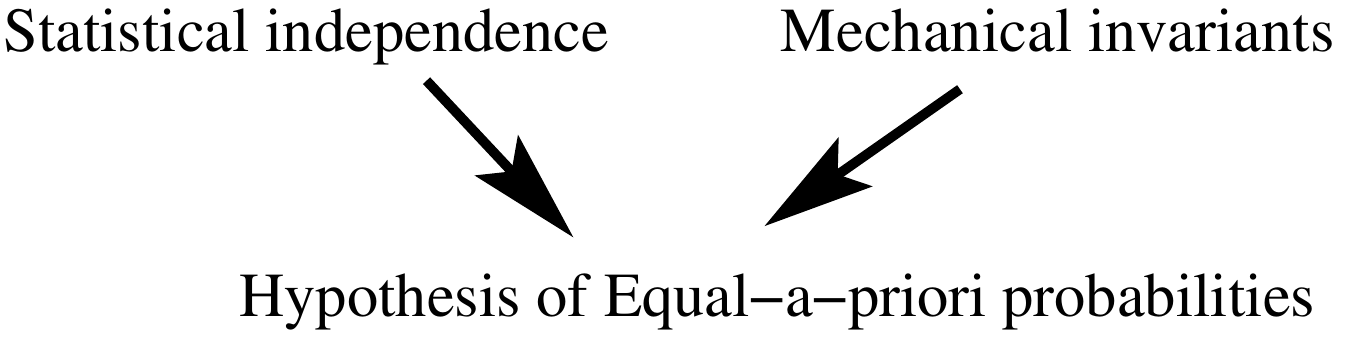}
\caption{The origin of equal-a-priori probabilities}
\end{figure}
As Tolman put it microstate has no preference to be in this or that part of phase space, but with the help of statistical independence it is more clearly seen. 

\nll
Reason (2): \footnote{Due to Prof. Sheldon Goldstein.} If one assume "statistical independence" then $\rho_{total} = \rho_{subsystem_1} \rho_{subsystem_2}.....\rho_{subsystem_N}$. Let $f_{total} = \sum_i f_i$, where $f_i = \ln \rho_{subsystem_i}$. Now it is easy to prove that $\frac{\sqrt{\la \Delta f_{total}^2\ra}}{\la f_{total}\ra} \propto \frac{1}{\sqrt{N}}$. Thus $f_{total}$ is almost constant on the energy surface $E_{total} = E_{subsystem_1} + E_{subsystem_2} +.....$ for $N \rta \infty$. Which again force us to say that $\rho_{total}$ take almost constant value on the energy surface--a statement of equal-a-priori probabilities. Thus we see that the hypothesis of statistical independence directly gives us equal-a-priori probabilities. 

The important point here to be noted is that the above reasons are only qualitative observations based on statistical independence and these are not ``rigorous" mathematical theorems. 

\begin{figure}
\includegraphics[height = 6cm, width = 5cm]{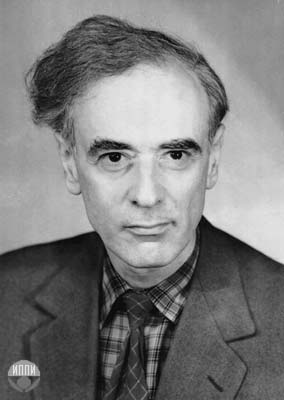}
\end{figure}

\subsubsection{Implications of Landau's argument: equilibrium statistical mechanics without any ad-hoc hypothesis and without the construction of ensembles}

Most fundamental ingredient: the property of "statistical independence".  

\nll
If one assume that various parts(macroscopic) of the body are statistically independent from each other, then one can construct canonical and microcanonical formulation without any ad-hoc and extra hypothesis (only based on statistical independence and some additive mechanical invariants of motion).  Consider a macroscopic system in equilibrium, all its parts (macroscopic) or subsystems of the whole system are in equilibrium with each other. Let $E_{subsystem}$ is the energy of the subsystem. It remains constant when the whole system is in equilibrium and the subsystem although small as compared to the whole system is still macroscopic and the relative fluctuations in $E_{subsystem}$ are very small $(\propto \frac{1}{\sqrt{N}})$.

One can argue that after a sufficiently long time intervals the subsystem cannot be considered quasi-closed,  as the effect of interaction of subsystems, however weak, will ultimately appear, and this weak interaction ultimately leads to the establishment of statistical equilibrium.  We will see below that in equilibrium this effect does not apply but in non-equilibrium it applies and we do not have any "non-equilibrium canonical" formulation.

Let us consider the subsystem for $\Delta t$ amount of time (much less than the relaxation time of the subsystem under consideration). The dynamics is Hamiltonian and due the Liouville's theporem $\log \rho_1$ is constant of motion. With Landau's argument $\log \rho_1 \propto E_{subsystem}$ or $\log\rho_1 = \alpha_1 + \beta_1 E_{subsystem}$. Now let us wait for a time interval sufficiently long as compared to the relaxation time. After this, the subsystem is no longer quasi-closed and say the distribution is $\rho_2$ at the end of this long time interval. Again consider an interval of time $\Delta t'$ much less than the relaxation time, during this interval system is again closed and Landau's argument again applies $\log \rho_2 \propto E_{subsystem}$ or $\log\rho_2 = \alpha_2 + \beta_2 E_{subsystem}$. From the above two equations we have $\alpha_1-\alpha_2 +(\beta_1 -\beta_2) E_{subsystem} = \log\frac{\rho_1}{\rho_2} $.  Now thermodynamic considerations show that $\beta$ is to be identified with inverse temperature of any subsystem of the whole system. As the whole system is in equilibrium thus all the subsystems of the whole system are in equilibrium with each other thus $\beta_1 = \beta_2$. Also normalization conditions demand that $e^{\alpha_1} = e^{\alpha_2} = \frac{1}{\int e^{\beta E_{subsystem}} dpdq}$. This gives us $\rho_1 = \rho_2$ and the distribution remains stationary. Thus in equilibrium Landau's argument applies at all times. Moreover $\rho = e^{\alpha + \beta E_{subsystem}}$ is constant, as RHS of this equation is constant both in time and phase (we neglect the fluctuations in $E_{subsystem}$). Thus $\rho = constant$, which is microcanonical ensemble.

On similar lines with Landau's argument we can proceed to show the canonical distribution $\rho = e^{\alpha + \beta E_{subsystem}}$. Here we allow fluctuations in $E_{subsystem}$ but keep $\beta$ same for all subsystems.

{\it The most important point to be noted here is that we do not "mentally construct" an ensemble and we do not assume equal-a-priori probability hypothesis. All the statistical formulation comes out from the property of ``statistical independence"    and the few additive integrals of motion (note that statistical independence does not apply to long range interacting systems). All other integrals of motion does not play any role in the Landau's argument because they are not additive.}

\subsubsection{Limitations of Landau-Lifshitz's approach}

(1)Temporal definition of probability: In a sufficiently long interval of time $T$ the phase point spends $\Delta t$ amount of time in a given volume $\Delta p \Delta q$ of the phase space, thus $ w = \lim_{T \rta \infty}\frac{\Delta t}{T}$ is the probability that, if the system is observed at any arbitrary time, its phase point will be found in the volume $\Delta p \Delta q$.

\vspb
In the above definition, non-stationary case cannot be defined, as the probability to find the system in $\Delta p\Delta q$ itself change with time. Thus the above definition excludes ubiquitous non-equilibrium cases.

\vspb
(2) As Landau-Lifshitz's whole analysis is based upon quasi-closed subsystems, the relation $  log(\rho) = \alpha + \beta E(p,q) + {\mathbf{\gamma . P}(p,q) + {\mathbf{\delta . M}}(p,q)  }$ holds good only for ``not too long intervals of time". As after a sufficiently long time intervals the subsystem cannot be considered quasi-closed. In their own words "Over a sufficiently long interval of time (compared to the relaxation time),  the effect of interaction of subsystems, however weak, will ultimately appear. Moreover, it is just this relatively weak interaction which leads finally to the establishment of statistical equilibrium". We have seen that in equilibrium Landau's argument applies at all times, but when the system is not settled to equilibrium (temperature equilibrium) we cannot apply Landau's argument due to the above reason.

\subsection{E. T. Jaynes' approach}

There are two schools at the foundations of probability theory (1) frequentists (probability from many random experiments), and (2) Bernoulli-Bayes-Laplacians or simply Bayesians (Principle of indifference). 

\vspb
Jaynes' standpoint: The ergodic problem is a conceptual problem related to the very concept of ensemble itself which is a byproduct of frequency theory of probability. Ergodic problem becomes devoid of any meaning when the probabilities of various micro-states are interpreted with Bernoulli-Bayes-Laplace theory of Probability.  In the Bayesian theory of probability, probabilities of occurrence of events are independent of the frequency concept. It is a more general viewpoint in which frequency theory is a special case and is based upon the principle of Indeference. Thus if we do not visualize the probability of occurrence of a micro-state with frequency (or construct ensemble) there is no question of ensemble averaging and ergodic problem is completely bypassed.

\vspb
Then the statistical mechanical theory is attacked with statistical inference (Shannan entropy approach). In Jaynes' words\cite{jaynes79}:
\nll
 "We can have our justification for the rules of statistical mechanics, in a way that is incomparably simpler than anyone had thought possible, if we are willing to pay the price. The price is simply that we must loosen the connections between probability and frequency, by returning to the original viewpoint of Bernoulli and Laplace. The only new feature is that their principle of Insufficient reason is now generalized to the principle of Maximum Entropy."

For more details see his detailed account\cite{jaynes79}.

\subsubsection{Critique of Jaynes approach}

The present author feels that it does not matter (atleast  in the computational problems of statistical mechanics) that which theory of probability one is accepting. But there is a disadvantage if we attack statistical mechanical problem with Bayesian viewpoint.

\vspb
If one accepts the Bayesian viewpoint, then one abandons the concept of ensembles. The probabilities of the microstates are obtained by maximizing the Shannon's entropy (with the given amount of macroscopic information). But this also means that one is neglecting the dynamical properties of constituents of matter (although they are not so important for sum-functions, a-la Khinchin). Obviously molecules of a gas obey Newton's laws or Schroedinger's equation. The phase point of an isolated system has no preference for this region or that region of phase space (the phase point of an isolated system visits all the accessible regions of phase space, {\it with equal frequency} although time involved are fantastically large). 

It is also the fact that in macroscopic observables (which are so coarse on microscopic scale) the microscopic dynamical properties do not reflect {\it in all its detail}, only very general microscopic dynamical properties reflect. For example, additive mechanical invariants play the essential role (in Landau's approach). Thus neglecting the dynamical properties of the constituents altogether is not a valid starting point. Thus it seems difficult to describe the integrable to non-integrable transition in dynamical systems with Jaynes approach. 

As we have seen that the sum-function macroscopic observables are so weakly depend on the dynamics of the phase point and thus this is the resolution (with Khinchin), not with the Bayesian viewpoint, which makes ergodic problem devoid of any meaning. Historically it is to be noted that Prof. G. Uhlenbeck objected Jaynes ideas\cite{jaynes79}.



\section{Quantum foundations of statistical mechanics}

Since in the quantum case position and momentum do not commute with each other, the classical concept of phase space that helps us to envisage the dynamics of the phase point on the energy surface breaks down.  However an equation analogous to equation (1) can be written in the quantum case also. The equivalent of "phase space energy shell" in the quantum case is the sub-Hilbert space $(\mathcal{H}_\nu)$ of the total Hilbert space $\mathcal{H}$. $\mathcal{H}_\nu$ is defined as the space spanned by all $\phi_\nu$ (eigenstates of the system's Hamiltonian) such that their corresponding eigenvalues $E_\nu$ are in a specified interval of energy from $E$ to $E + \Delta E$. The quantum equivalent of the classical ergodic hypothesis (equation (1)) is written as 

\begin{equation}
\lim_{T\rightarrow \infty} \int_0^T \la A(t) \ra dt = \sum_{\alpha : E_{\alpha}\in E, E + \Delta E } |c_\alpha|^2 \la \phi_\alpha |\hat{A}|\phi_\alpha \ra,
\end{equation}

for a system in pure quantum state (for details see equation (11) below). For a system in mixed quantum state replace $|c_\alpha|^2 \rightarrow \rho_{\alpha\alpha}$ with $\rho_{\alpha \alpha}$ as the diagonal element of the density matix in the energy representation.

{\it Here again we encounter (LHS of the above equation) the problem of infinite time averaging, and all the arguments of section I(A) are valid, but we will see below that the "resolving results" analogous to the Khinchin's self averaging theorem hold in quantum case also--either based upon quantum chaos or based on von Neumann's "scale separation" ideas. We again see the fundamental role of typicality and large dimensional Hilbert spaces}.

\subsection{Eigenstate Thermalization Hypothesis (ETH)}

Eigenstate Thermalization Hypothesis(ETH) implies that the thermalization happens at the level of individual eigenstates. Consider a system with Hamiltonian $H$ and eigensystem $E_\alpha,~~\phi_\alpha$ prepared in some initial state. If system has unitary evolution, then at any later time $t$

\[|\psi(t)\ra = \sum_{\alpha} c_\alpha e^{-i E_\alpha t/\hbar}  |\phi_\alpha\ra\]

Quantum mechanical mean of an operator $\hat{A}$ pertaining to the system is:

\[\la \hat{A}(t)\ra = \la \psi(t) |\hat{A}| \psi(t)\ra = \sum_{\alpha,\beta} c_{\alpha}^\ast c_{\beta} e^{i(E_{\alpha} -E_\beta)t} \la \phi_\alpha |\hat{A}|\phi_\beta \ra\]

Consider infinite time average

\begin{eqnarray}
\overline{\la \hat{A}(t) \ra} = &&\sum_{\alpha,\beta} c_\alpha^\ast c_\beta \delta_{\alpha,\beta} A_{\alpha,\beta} = \sum_{\alpha} |c_\alpha|^2 A_{\alpha,\alpha}~~,\nonumber\\
&&\left(\delta_{\alpha,\beta} = \lim_{T\rightarrow \infty} \frac{1}{T}\int_0^T e^{i(E_\alpha -E_\beta)t} dt\right)
\end{eqnarray}

$\sum_{\alpha} |c_\alpha|^2 A_{\alpha \alpha}$ is also called the diagonal ensemble in\cite{rigolnature}. Now the thermodynamic universality demands:

\vspb
LONG TIME AVERAGE $\equiv$ AVERAGE OVER APPROPRIATE STATISTICAL ENSEMBLE (MICROCANONICAL OR CANONICAL ETC.)

\begin{eqnarray}
&&\sum_\alpha |c_\alpha|^2 A_{\alpha,\alpha} = \la A\ra_{microcan~at~E=E_0}\nonumber\\
&&=\frac{1}{N_{E_0,\Delta E}} \sum_{\alpha; |E_0-E_\alpha| < \Delta E} A_{\alpha,\alpha}
\end{eqnarray}

This is an universality relation with LHS depends on initial conditions but RHS does not !

{\it Now Eigenstate Thermalization Hypothesis (ETH) says that there are no  eigenstate--to--eigenstate fluctuations in  $A_{\alpha,\alpha}$'s for eigenstates close in energy}
\[\Rightarrow A_{\alpha,\alpha} \sum_{\alpha \in Window}|c_{\alpha}|^2  = A_{\alpha,\alpha} = \frac{1}{N_{E_0,\Delta E}} \sum_{\alpha\in Window} A_{\alpha,\alpha}\]

This is called the eigenstate thermalization hypothesis (ETH) by Deutsch (1991) and Srednicki (1994)\cite{deutsch,sred94}:
\nll
"The quantum expectation value $\la \psi_\alpha |\hat{A}|\psi_\alpha\ra$ in a large interacting many-body system is equal to the thermal average"

\begin{equation}
\boxed{\la \psi_\alpha |\hat{A}|\psi_\alpha\ra = \la A \ra_{microcan. ~around ~~E_\alpha}}
\end{equation}

\begin{figure}
\includegraphics[width= 9cm, height=7cm]{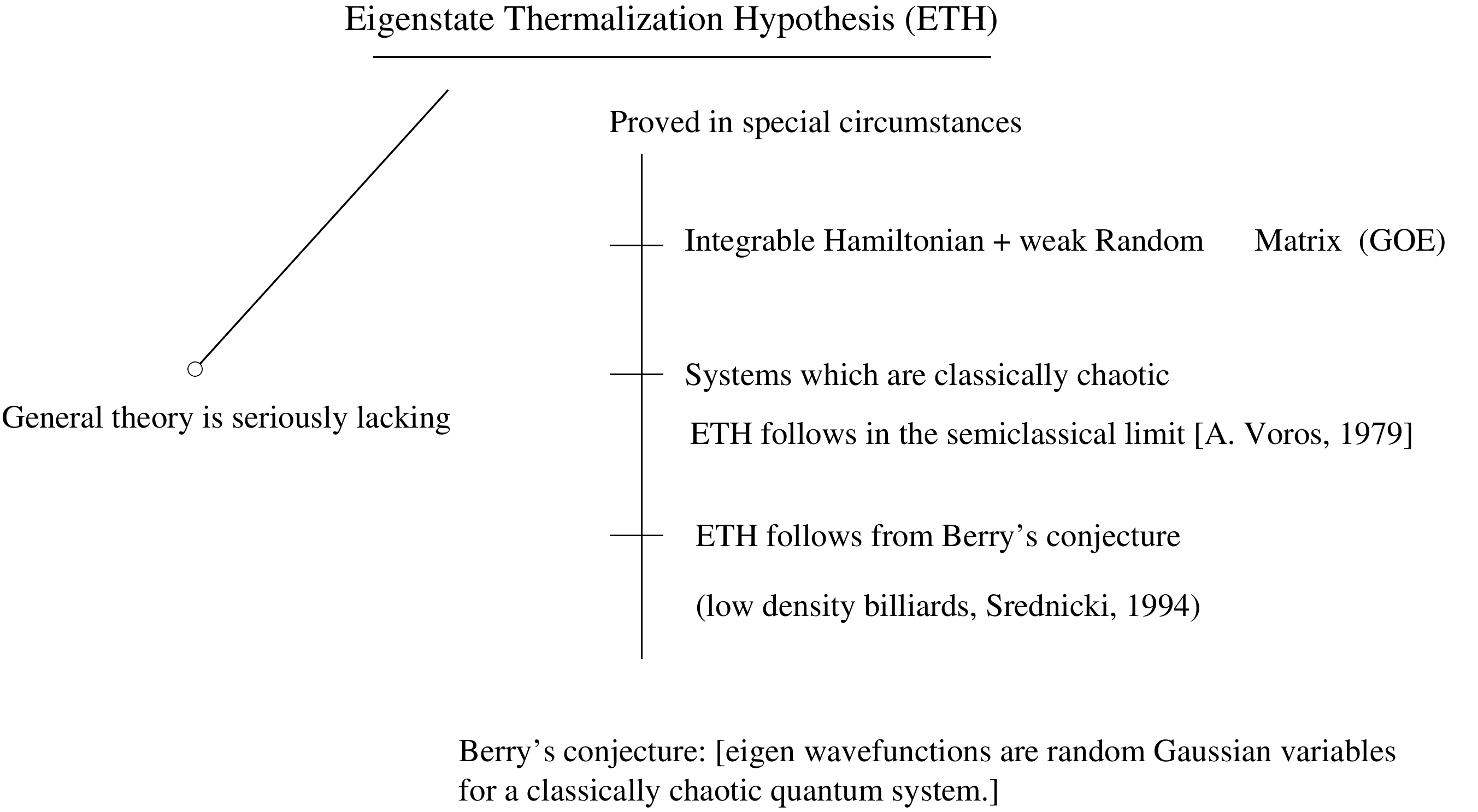}
\caption{Status of ETH}
\end{figure}

If this is valid in generality, then one explain thermodynamic universality. 
Also if the system obeys Berry's conjecture (eigenfunctions are random Gaussian variables for a classically chaotic quantum system) the off-diagonal elements $(\la \psi_\alpha |\hat{A}|\psi_\beta\ra,~~\alpha\ne\beta)$ are negligible and {\it then one obtains thermal behaviour without any time averaging at all}\cite{sred94} . Also it has been numerically shown\cite{rigolnature} that the magnitude of the off-diagonal elements of the momentum distribution operator are very small as compared to the diagonal elements $A_{\alpha,\beta}|_{\alpha\ne\beta} << A_{\alpha\alpha}$. Clearly much work is needed to show ETH for other operators and {\it  importantly for sum-function operators} (across the integrability-to-(non)integrability transition). For detailed discussion see\cite{rigolnature} and the ref\cite{rigol2009a,rigol2009b} for the behaviour of diagonal and off-diagonal elements of various other observables.

Study of literature shows that ETH has been proved analytically only in special circumstances (see figure  5). {\it Figure shows that some kind of chaos is necessary for ETH to hold}. This is also evident in the numerical experiment of Rigol etal\cite{rigolnature} for a small system (five hard core Bosons on a lattice) that ETH hold good in the non-integrable system but fails in the integrable one (for marginal momentum distribution as the observable).

Clearly systems with very small number of degrees-of-freedom cannot be fit into Khinchin's program (which requires macroscopic systems). This implies that the statistical mechanical universality (irrelevance of chaos) that we enjoy with macroscopic systems and sum function observables may no longer holds good for small isolated quantum systems.

\subsection{von Neumann's quantum ergodic theorem (or the better notion of "normality" due to Goldstein-Lebowitz-Tumulka-Zanghi (GLTZ))}

We present here the qualitative statement of von Neumann's quantum ergodic theorem, for a precise definition see\cite{gltz}. Our aim here is to understand the physical basis of the theorem,  to draw some analogies with the approach by Khinchin,  and to see the connection with the ETH (eigenstate thermalization hypothesis).

Setup: Consider a system with Hamiltonian $H$ and total Hilbert space $\mathcal{H}$. Consider that the total Hilbert space is partitioned  into mutually orthogonal sub-Hilbert-spaces $\mathcal{H}_\nu$, with family $\mathcal{D}$ ($\mathcal{D} \equiv \{\mathcal{H}_\nu\},~~~\mathcal{H} = \bigoplus \mathcal{H}_\nu$). Let $d_\nu = dim \mathcal{H}_\nu$ and $D= dim \mathcal{H}$. Let total number of partitions is $n$ i.e., $ n = \# \mathcal{D}$.

The physical basis of this partitioning (according to von Neumann) is that each $\mathcal{H}_\nu$ belongs to a different macro-state (or macro-observer). This is the crucial insight of von Neumann, as in quantum theory all operators corresponding to observables do not commute among themselves. von Neumann take a physical stand point that the measurement process is coarse on microscopic variables, thus operators can be taken "approximately" commuting (measurement errors are large as compared to bounds of uncertainty relations). This is called "rounding" of operators.

The aim of the Quantum Ergodic Theorem (QET) is to tell us some kind of universality or "normality" that the quantum expectation values of operators are close to the microcanonical averaging for most of the time (under some conditions, see below). It is a kind of quantum H-theorem which says that for all initial states the time evolution take the system from initial non-equilibrium state to the equilibrium state (provided the system satisfy some conditions).

Thus the total Hilbert space is partitioned in sub-spaces $\mathcal{H}_\nu$ and the macroscopic observables take specific values in each $\mathcal{H}_\nu$ (values taken in $\mathcal{H}_\nu$ are different from those taken in $\mathcal{H}_\mu$ when $\nu\ne \mu$). The "rounded" operators commute with each other but they do not commute with the Hamiltonian (this is analogous to non-integrals of motion in classical Gibbs picture).

\vspb
Qualitative statement of QET:

Let $P_\nu$ be the projection to $\mathcal{H}_\nu$ with macro-state $\nu$ having some values of observables.
\nll
Then QET says:

For {\it all} initial wavefunctions  $\psi_0 \in \mathcal{H}, ~~ ||\psi_0|| =1$ (note that this includes the  special initial non-equilibrium states), we have for most of the time

\begin{equation}
\boxed{||P_\nu \psi_t||^2 = \la \psi_t |P_\nu|\psi_t\ra \simeq tr(\rho_{mc} P_\nu) = \frac{d_\nu}{D}}
\end{equation}

In words: probability distribution over macro-states is approximately equal to the  ratio of the dimension of Hilbert space corresponding to that macro-state to the dimension of the total Hilbert space. Thus the macro-state with largest $d_{\nu}$ is most probable  (which is in accord with common sense, equilibrium state has the largest $ d_\nu = d_{eq}$ and it is typically observed).

Conditions involved in QET are:

\begin{enumerate}

\item Hamiltonian is free of resonances i.e., $E_\alpha -E_\beta \ne E_{\alpha'} -E_{\beta'}$, unless either $\alpha = \alpha',~~~\beta = \beta'$ or $\alpha = \beta,~~~\alpha' =\beta'$.

\item The quantity $f_\nu(H,\mathcal{D}) = max_{\alpha\ne\beta} |\la \phi_\alpha|P_\nu|\phi_\beta\ra|^2 + max_{\alpha} (\la\phi_\alpha|P_\nu|\phi_\alpha\ra -\frac{d_\nu}{D})^2$ is small for all $\nu$. The meaning of small here is that the quantity $f_\nu(H,\mathcal{D})$ is smaller than $\ep^2\frac{d_\nu}{n D}\frac{\delta'}{n}$. Here $\ep$ and $\delta'$ are small positive numbers and $n$ is the number of partitions $n = \#\mathcal{D}$. For more details see\cite{gltz} and English translation of the original von Neumann's article\cite{neumann}.

\end{enumerate}

This is QET. This property should be better called "normality" after GLTZ. GLTZ also give a stronger bound on the deviations from the average\cite{normal}. The notion of "normality" is a special case of a broader notion of "typicality". There are various kinds of normality properties as discussed in the next subsection. Strict determinism of classical mechanics is replaced by the notion of "typicality" of statistical mechanical systems. The notion of typicality captures the {\it almost} deterministic behaviour of statistical mechanical systems (although atypical behaviour is possible in principle but highly improbable)
\footnote{Dictionary meaning of typicality is: serving as a type  or representative specimen or conforming to a particular type. For example, typical lottery tickets are empty, typically monsoon comes in the months of May to August in India, a gas initially enclosed in the left half of the box "typically" expands to occupy the full volume available when the partition is removed (it is highly improbable that the gas will come back to the left half again in the terrestrial time scales, {\it but it is not impossible}).}.

\subsubsection{Remarks on ETH and QET}

\begin{enumerate}


\item  QET is a typicality result (typicality due to "scale separation"(see DNT below)) valid for large systems (i.e., containing large number of particles) but ETH (again a typicality result--typicality due to quantum chaos) has been applied even for few boson atoms in an optical lattice.

\item ETH involves quantum chaos and QET does not involve quantum chaos !

\item ETH has been proved (analytically) only in special circumstances based on the assumption of quantum chaos\cite{sred94}, see figure(5). General theory is seriously lacking.

\end{enumerate}

\begin{figure*}
\includegraphics[width= 18cm, height=10cm]{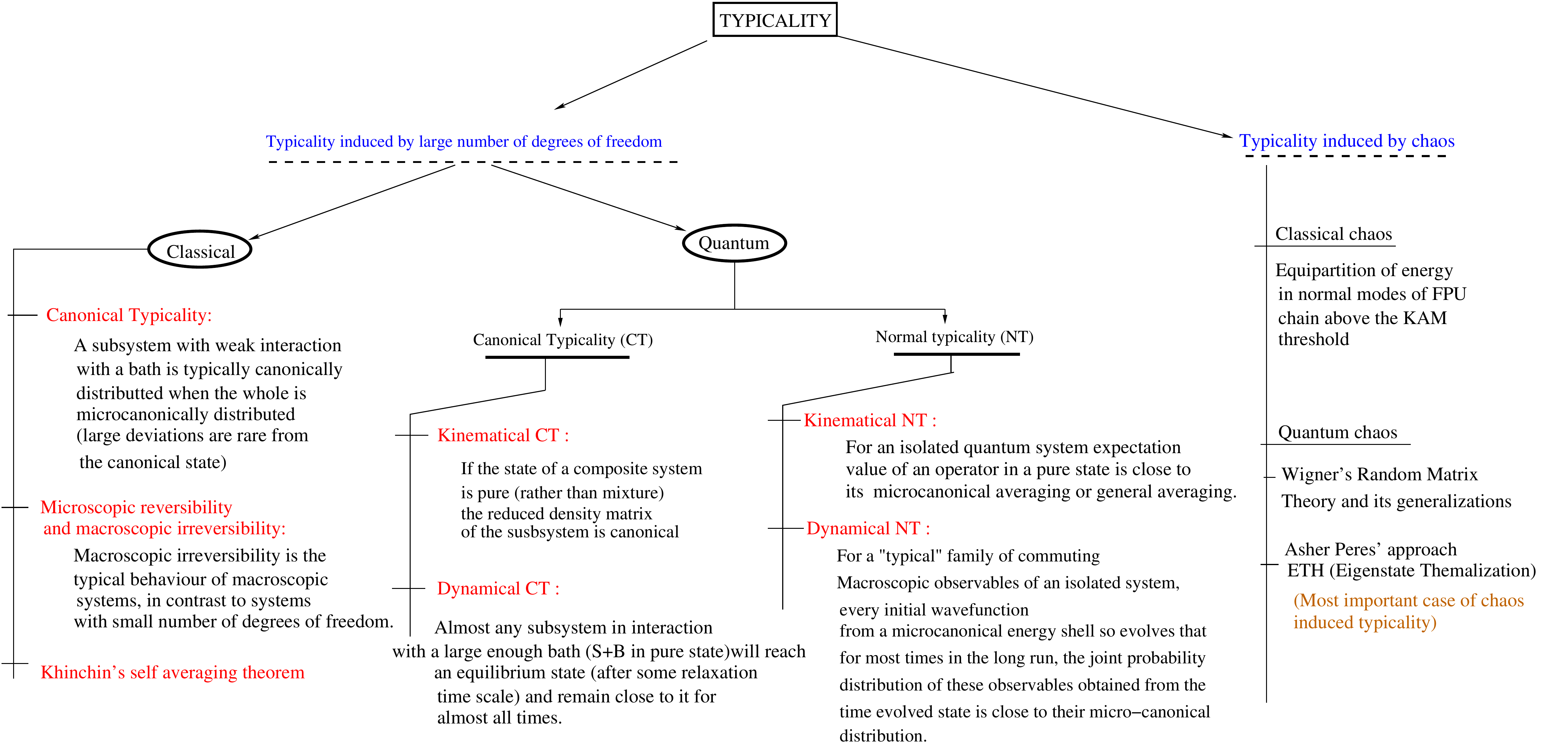}
\caption{Typicality of statistical mechanical systems}
\end{figure*}

\subsection{Other approaches}

Recent literature can be organized under the following four notions (see figure 6):

\subsubsection{Kinematical canonical typicality (KCT)}

Qualitative definition: We know that a small system weakly coupled to a large bath is described by canonical ensemble when the composite system (system + bath) is described by microcanonical ensemble.  Kinematical canonical typicality says that even if the state of a composite system is pure (rather than a mixture as in the traditional case) the reduced density matrix of the system is canonical.

KCT has been announced in\cite{sheldon2006} extending the works of Schr\"{o}dinger\cite{s52}. Related works appeared in\cite{mahler, tasaki}. In \cite{sheldon2006} the basic assumption is that the probability distribution is uniform over all normalized wavefunctions $\Psi$ with energy in the  shell $[E,E+\delta]$ of the composite system. The system density matrix (when the composite is in pure state) $\rho^{\Psi} = tr^{B}|\Psi\ra\la \Psi|$ is obtained by tracing out bath degrees of freedom. The crucial role in the tracing is played by the expansion co-efficients $C_{ij}$ in $\Psi=\frac{\sum_{i,j} C_{ij} | E^{S}_i\ra | E_j^{B}\ra}{||\sum_{i,j} C_{ij} | E^{S}_i\ra | E_j^{B}\ra ||}$ which were assumed Gaussian Random variables. It is shown that $\rho^{\Psi}$ is approximately canonical (see for details\cite{sheldon2006}). A rather general treatment of KCT is given in\cite{sandu}. They prove KCT using Levy's Lemma which is analogous to the law of large numbers. They prove that $\la D(\rho^{\Psi}, \Omega_S)\ra \leq \frac{1}{2}\sqrt{\frac{d_S}{d_E^{eff}}}$, where $D(\rho^{\Psi}, \Omega_S)$ is the trace distance $\frac{1}{2} tr \sqrt{(\rho^{\Psi} -\Omega_S)^\dagger(\rho^{\Psi} -\Omega_S)}$ between $\rho^{\Psi}$ (reduced density matrix of the system when the composite is in the pure state) and $\Omega_S$ (the traditional (when composite is in the mixed state) reduced density matrix of the system). The average $\la ...\ra$ is over the environment states with the standard unitarily invariant measure. $d_S$ is the dimension of the system's Hilbert space and $d_E^{eff}= 1/tr(\Omega_E^2)$ is a measure of the effictive size of the environment $d_E^{eff} \geq \frac{d_R}{d_S}$ ($d_R$ is the dimension of the environment's Hilbert space). Thus when bath or environment is much larger than the system i.e., $d_R>>d_S$ then $d_E^{eff}>>1$ and if $d_S/d_E^{eff}$ is very small then $\rho^{\Psi}$ and $\Omega_S$ are close to each other, $\la D(\rho^{\Psi}, \Omega_S)\ra \leq \frac{1}{2}\sqrt{\frac{d_S}{d_E^{eff}}}$, which is a statement of KCT (for details and related results see\cite{sandu}).

\subsubsection{Dynamical canonical typicality (DCT)}

Qualitative Definition: When the composite system (system + Bath) is in a pure state, the expectation value of an operator pertaining to the system, {\it after sufficiently large time}, will be almost equal to the canonical expectation value.

DCT has been proved in\cite{tasaki} and there were numerical experiments\cite{nexp} for its evidence. In \cite{tasaki} DCT has been proved with an assumption of weak coupling between system and bath with $\Delta\ep>>\lambda>>\Delta B$ ($\Delta\ep$ is the minimum spacing between the energy levels of the system, $\lambda$ is the magnitude of system-bath coupling, and $\Delta B$ is the maximum spacing between the energy levels of the bath). Also a hypothesis of "equal weights for eigenstates" is used which makes the energy expansion co-efficients small. The statement of DCT goes like this. Suppose that the composite-system is in the pure state $|\Psi(t)\ra$ at any time $t$, the expectation value of an operator of the system is written as $\la A\ra_t = \la \Psi(t)| (A  \otimes I_{B}) |\Psi(t)\ra $. Then DCT says, after sufficiently long time, $\la A\ra_t \simeq \frac{tr_S(A e^{-\beta H_S})}{Z}$. For   proof see\cite{tasaki}. DCT has been proved in a much more general setting in\cite{lindenpre79} and  also in\cite{cho} using an assumption that matrix elements of the interaction Hamiltonian has random phases and then constructing Markovian master equations. In \cite{lindenpre79},  the authors extend their previous kinematical results (last subsection). They prove that $\la D[\rho_{S}(t),\omega_S]\ra_t \leq \frac{1}{2}\sqrt{\frac{d_S}{d^{eff}(\omega_B)}} \leq  \frac{1}{2}\sqrt{\frac{d_S^2}{d^{eff}(\omega)}} $ . Here $\rho_S(t) = tr_B[ \rho_{total}(t)]$ is the system density matix ( $\rho_{total}(t) = |\Psi(t)\ra\la \Psi(t)|$ is the total density matrix (system + Bath)) and $\omega_S \equiv \la \rho_S(t)\ra_t = \lim_{T\rightarrow \infty} \int_0^T \rho_{S}(t) dt$. In words, theorem says that the time average of the trace distance (see previous subsection) between instantaneous system's density matrix and its time averaged density matrix is small if $ d_S << d^{eff}(\omega_B)$. This means that system spends almost all of its time near to $\omega_S$. This is called equilibration (weakly fluctuating about a steady state---this steady state need not be an equilibrium state). The important point is that this result is completely general (except the assumption that the Hamiltonian is free of resonances--condition no (1) in von Neumann's theorem). It does not depend upon the nature of system-bath coupling, condition of bath (whether it is in equilibrium or not). They also treat the problem of thermalization (i.e., initial state independence, as the steady state reached may depend on the initial conditions). For a detailed account and further results, see their very nice paper\cite{lindenpre79}.  Our purpose here is to introduce various typicality theorems and to give the excitement to the reader. Interested reader should go to the original literature.

\subsubsection{Kinematical Normal typicality (KNT)}

Qualitative Definition: For an isolated quantum system in an arbitrary pure state, the quantum expectation value of an operator in that state {\it hardly} deviates from the ensemble average. The meaning of "hardly" is clarified below.

A clear account of KNT is given in\cite{reimann}. The statement of KNT theorem there  goes like this; Consider that the system is in some pure state $|\Psi \ra = \sum\frac{c_n}{\sqrt{\sum |c_n|^2}} |n\ra$, with $|n\ra$ as eigenstates. The quantity $\sigma_A^2 \equiv \overline{(\la \Psi |A|\Psi\ra - \overline{\la \Psi|A|\Psi \ra})^2}$ is small. Here the average $\overline{\la\Psi| A|\Psi \ra} \equiv \int \la \Psi |A|\Psi\ra p(c) \prod_{n=1}^N d Re(c_n) d Im(c_n)$, where $p(c)$ is the probability distribution over the expansion co-efficients. The meaning of "small" in the above statement is that,
$\sigma^2_A\leq \Delta_A^2 (max q_n) (tr \rho^2)$, here $q_n$'s are positive numbers typically of order one, and $\Delta_A$ is the difference between the maximum and minimum eigenvalues of $A$. For $K$ operators one has  $  Probability[max_{i} \frac{|\la \Psi |A_i|\Psi\ra - \overline{\la \Psi|A|\Psi \ra} |}{\Delta_i} \geq \ep] \leq \frac{K (max_{n} q_n) (tr \rho^2)}{\ep}$. Where for $K$ operators $i$ runs from $1$ to $K$, and $\Delta_i$ is the difference between the maximum and minimum eigenvalues of $A_i$.

The smallness of $\sigma_A^2$ (which is must for typicality) is imposed due to the conditions involved in the above theorem;

\begin{enumerate}
\item The expansion co-efficients $c_n$ are statistically independent and $c_n,~~c_n^{i \phi_n}$ are equally likely for arbitrary phase $\phi_n$. Thus $p(c) = \prod_i^N p_n(|c_n|)$.

\item The mixed state $\rho = \overline{|\Psi \ra\la \Psi|} =\sum\overline{(\frac{|c_n|^2}{\sum |c_n|^2})} |n\ra \la n|$, with (overbar means average over $p(c))$ has low purity $ tr \rho^2 <<1$.

\end{enumerate}

The author\cite{reimann} justify these assumptions in a qualitative  way that these are valid in "practice" if not "in principle" for a system with large number of degrees of freedom. The rigorous justification is serious lacking. The role of quantum chaos in these theorems is not clear at present. KNT is also discussed in \cite{hal} where it is termed as "thermodynamic normality"

\subsubsection{Dynamical Normal typicality (DNT)}

The von Neumann's quantum ergodic theorem is the first known theorem of dynamical normal typicality. This theorem is based upon the observer's inability to further resolve $\mathcal{H}_\nu$ by macroscopic means and the bound on fluctuations of the quantum expectation values of the operators from microcanonical averaging due to the structure imposed on the Hilbert space by macroscopic perspective (the large $D = dim \mathcal{H}$ and large $n = \# \mathcal{D}$, see subsection B). 

Important point is that {\it no} chaos or disorder assumption was invoked by Neumann. In his own words\cite{neumann} "...we emphasize that the true state (about which we do calculations) is a wavefunction, i.e., something microscopic---to introduce a macroscopic description of the state would mean to introduce disorder assumptions, which is what we definitely want to avoid". {\it His work is a result of "scale separation" ( i.e., observer's inability to further resolve $\mathcal{H}_\nu$ by macroscopic means, the large $D = dim \mathcal{H}$ and large $n = \# \mathcal{D}$ etc)}. Spirit of his work resembles with that of Khinchin and  more recent works\cite{sandu,dntreimann}, in contrast to ETH  which involves quantum chaos.

This state of affairs (un-necessity  or necessity of disorder) make the situation complicated and no satisfactory resolution is available at present.  We again consider the role of quantum chaos (previous example was ETH).

The important work in this direction (quantum chaos in the foundations of statistical mechanics) was done by Asher Peres in 1980's\cite{peres}. He discusses DNT based upon his definition of quantum chaos that the observables are represented by pseudorandom matrices when the Hamiltonian is diagonal. Due to this, expectation values of these observables tend to equilibrium values and fluctuations around these equilibrium values are, on the average, very small. His argument for quantum ergodicity goes like this; the quantum expectation value of an observable is $\la A(t) \ra = \sum_{E' E"} \rho_{E' E"} \la E'|A| E"\ra e^{i t (E'-E'')}$ for non-degenerate spectrum, time average is $\overline{\la A(t) \ra} = \sum_E \rho_{EE} A_{EE} $ (here $\rho_{EE}$ is the density matrix in energy representation). Here comes the Peres' argument; consider that the system has very many energy levels in the range $E$ to $E +\Delta E$, (1) distribution of these energy levels is very very different in regular and chaotic case\cite{gut,balian,berry}, (2) the observable $A_{EE}$ too behave very differently for regular and chaotic case  (regular system have selection rules, most $A_{EE}$ vanish and only a few are large, and in chaotic case $A_{EE}$ is pseudorandom (his definition of quantum chaos) and very numerous), and (3) for a chaotic system, the pseudorandom $A_{EE}$ are {\it statistically independent} from $\rho_{EE}$ in that energy range and {\it their average does not appreciably depend on the energy interval $\Delta E$}. As $\sum_E\rho_{EE} =1$, one has $\overline{\la A(t)\ra} = \bar{A}(\bar{E})$. This he calls quantum ergodicity. The fundamental assumption used is the assumption of statistical independence of $A_{EE}$ from $\rho_{EE}$.  Then he discuss more relevant property, called mixing in quantum case,  that the fluctuations of $\la A(t)\ra$ about its equilibrium value $\bar{A}(\bar{E})$ are, on the average,  small i.e., $\sqrt{\overline{\la A(t)\ra^2} - (\overline{\la A(t)\ra})^2} \lesssim\sqrt{\frac{\bar{A^2}(\bar{E})}{N}}$ ($~~N$ is the large number of energy levels in the interval $E$ to $E + \Delta E$). To prove this he again used the assumption of statistical independence. Thus we see from Peres' program that DNT is a consequence of quantum chaos (observables as pseudorandom matrices) and {\it the assumption of statistical independence.}

This is in sharp contrast with the approach of von Neumann (as von Neumann did not use quantum chaos), still we do not have a coherent picture that when chaos is important (Peres) and when chaos is not important (von Neumann)? Various typicality approaches are summarized in figure (6).

\section{Summary and open issues}
\begin{enumerate}

\item Justification of Gibbs' ensembles is a complicated problem.

\item Ergodic hypothesis is not necessary for the workings of statistical mechanics as far one is concerned with sum-function observables and macroscopic systems.

\item Statistical mechanics explain thermodynamic behaviour because of the property of statistical independence, or, more accurate approach of Khinchin which is the cause of typicality and canonical formalism can be obtained from Landau's program without using any ad-hoc hypothesis except the property of statistical independence.

\item The issues of chaos, integrability or non-integrability becomes important when one deals with small isolated systems, for example, $100$ or so Bosonic or Fermionic atoms in an optical lattice. They are irrelevant for a macroscopic system and sum-function observables.

\item Clearly von Neumann's quantum ergodic theorem and other quantum-statistical typicality theorems are analogous to  Khinchin's self averaging theorem. These become more accurate for a system with large number of degrees of freedom and these say that for such systems the expectation values of the observables (sum-functions in Khinchin's case) stay close to the microcanoncal/canonical predictions. In principle,  Khinchin's approach should come out as a special case of von Neumann's theory. Clearly much work is needed in this direction.

\item One can contemplate {\it classical} KCT (this traditional statistical mechanics), {\it classical} DCT, and so on.

\item The above presented consolidation or amalgamation of various quantum-statistical typicality theorems singles out  ETH and Peres' DNT.  ETH and Peres' DNT involves quantum chaos, whereas others like von Neumann's do not involve quantum chaos. Still we do not have a coherent picture; when chaos is important (Peres) and when chaos is not important (von Neumann)?

\end{enumerate}

I would like to end this manuscript with the following aphoristic words of Prof. Jeol Lebowitz\footnote{Vienna International Symposium in Honour of Boltzmann's 150-th birthday, 1994 (page 13, R. L. Dobrushin, "A mathematical approach to the foundations of statistical mechanics" Vienna, Preprint ESI 179 (1994))} about the ergodic hypothesis;
\nll
"Now it is the real time to recognize that the Ergodic Hypothesis is not a necessary and is not a sufficient condition for the foundation of statistical mechanics"

\acknowledgements

Author would like to thank to Prof. Sheldon Goldstein and Prof. Marcos Rigol for many critical discussions, to Prof. Angelo Vulpiani and Dr. Jitesh Bhatt for pointing out many corrections and suggestions. He is also thankful to Prof. Michael Kastner for giving him the opportunity to present this work at 2nd Stellenbosch Workshop on Statistical Physics from 7th  March to 18th March 2011 at Stellenbosch, South Africa,  and also for the partial financial help.


\section{Appendex}

\subsection{Boltzmann's permutational argument and the origin of $ S \propto ln W $}

Boltzmann introduced the notion of "Komplexions" of molecules as follows. Consider the kinetic energies of the molecules take the following discrete values $\ep,2\ep,3\ep,....n\ep$, and let $w_j$ number of molecules posses a kinetic energy $j\ep$. $\sum_{j=1}^n w_j = N,~~~~\sum_{j=1}^n (j\ep) w_j = E$. Number of Komplexions for a given distribution is $P = N!/w_1 ! w_2 !....w_n !$. He noted that maximization of $P$ with the above constraints leads to Maxwellian kinetic energy distribution.
\nll
Boltzmann then defines "degree of permutability" $\Omega = log P$, and finds that (by direct computation) that $\Omega_{max}$ is equal to the entropy of an ideal gas in a reversible process up to an additive constant.
\[S = \int \frac{\delta Q}{T} = \Omega_{max}= max[log P].\]

It was Max Planck who gave the general foundations to the fundamental principle $S = k_B log W + c$, (using $W$ in place of $P$) introducing first time the constant $k_B$ called the Boltzmann constant. He proposed the following fundamental hypothesis called Planck's thermodynamic probability hypothesis.


"The entropy of physical system in a definite state depends solely on the thermodynamic probability $W$ of this state", $S = f(W)$.
\nll
Considering two systems in equilibrium with each other. From second law, one has additivity of entropy $S = S_1 + S_2$, thus  $f(W_1 W_2) = f(W) = f(W_1) + f(W_2)$ and then by differentiating one can obtain $\dot{f}(W) + W \ddot{f}(W)=0$. It gives \[\boxed{S = k_B~ log W + c.}\]

See the little very good book by Enrico Fermi\cite{fermi}.

The important point here is that the equilibrium is conceived as the most probable state rather than the temporally constant state. But as we know that the fluctuations in observables are extremely small for macroscopic systems both definitions are approximately consistent with each other.


\begin{thebibliography}{201}

\expandafter\ifx\csname natexlab\endcsname\relax\def\natexlab#1{#1}\fi
\expandafter\ifx\csname bibnamefont\endcsname\relax
  \def\bibnamefont#1{#1}\fi
\expandafter\ifx\csname bibfnamefont\endcsname\relax
  \def\bibfnamefont#1{#1}\fi
\expandafter\ifx\csname citenamefont\endcsname\relax
  \def\citenamefont#1{#1}\fi
\expandafter\ifx\csname url\endcsname\relax
  \def\url#1{\texttt{#1}}\fi
\expandafter\ifx\csname urlprefix\endcsname\relax\def\urlprefix{URL }\fi
\providecommand{\bibinfo}[2]{#2}
\providecommand{\eprint}[2][]{\url{#2}}


\bibitem{Broglie} Page ix in the preface to the book "Foundations of classical and quantum statistical mechanics" by R. Jancel, Pergmon press Oxford, 1969.




\bibitem{Zubarev} Page 18 of his book "Nonequilibrium Statistical Thermodynamics", Consultants Bureau, New York, 1974.



\bibitem{JEMayer} Page 122 of their book "Statistical Mechanics", John Wiley and Sons, New York, 1977.


\bibitem{WTGrandy} Page xiii of his book "Foundations of Statistical Mechanics: Equilibrium Theory", D. Reidel Publishing Company, Holland, 1987.


\bibitem{ETJaynes1}Page 19, Where do we stand on Entropy?, presented at Maximum Entropy Formalism Conference, MIT, May 2-4, 1978.


\bibitem[{\citenamefont{Jeol Lebowitz}(1993)}]{lebowitz}
\bibinfo{author}{\bibnamefont{Lebowitz}, \bibfnamefont{Jeol}},
  \bibinfo{year}{2008}, {``}\bibinfo{title}{From time-symmetric microscopic dynamics to time-asymmetric macroscopic behavior: An Overview}{''},
  \bibinfo{journal}{in G. Gallavotti, W. L. Reiter, J. Yngvason (editors): Boltzmann's Legacy. Zurich: E. Math. Soc. },
  \bibinfo{pages}{63-88}.




\bibitem{yukalov} V. I. Yukalov, "Irreversibility of time for quasi-isolated systems", Phys. Lett. A {\bf 308}, 313 (2003).


\bibitem[{\citenamefont{Landau and Lifshitz}(2004)}]{LandauLifshitz}
\bibinfo{author}{\bibnamefont{Landau}, \bibfnamefont{Lev}}, and
  \bibinfo{author}{\bibfnamefont{Lifshitz}~\bibnamefont{E. M.}},
  \bibinfo{year}{1958}, {``}\bibinfo{title}{{\it Statistical Physics}}{''}, \bibinfo{journal}{Pregmon Press.}


\bibitem{jaynesL4}E. T. Jaynes, "Probability theory in Science  and  in Engineering", No. 4, in "Colloquium lectures in pure and applied science", Socony-Mobiloil co. USA, 1956.



\bibitem[{\citenamefont{Tolman}(1938)}]{Tolman}
\bibinfo{author}{\bibnamefont{Tolman}, \bibfnamefont{R. C.}},
  \bibinfo{year}{1938}, {``}\bibinfo{title}{{\it Principles of statistical mechanics}},{''} \bibinfo{journal}{Oxford university press}.




\bibitem[{\citenamefont{Anderson} (1972)}]{anderson}
\bibinfo{author}{\bibnamefont{Anderson}, \bibfnamefont{P. W.}},
\bibinfo{year}{1972}, {``}\bibinfo{title}{More is different}{''}, \bibinfo{journal}{Science}, {\bf 177}, \bibinfo{pages}{393}.



\bibitem[{\citenamefont{Ehrenfests}(1911)}]{ehrenfests}
\bibinfo{author}{\bibnamefont{Ehrenfest}, \bibfnamefont{Paul}}, and
\bibinfo{author}{\bibnamefont{Ehrenfest}, \bibfnamefont{Tatiana}},
  \bibinfo{year}{1990}, {``}\bibinfo{title}{The conceptual foundations of the statistical approach in mechanics},{''} \bibinfo{journal}{Dover publications, Mineola, N.Y.}




\bibitem{brush} Stephan G. Brush, {\it "The kind of motion we call heat"}, Vol.1 and 2., North-Holland, New York, 1976.


\bibitem{cer} Carlo Cercignani, {\it "Ludwig Boltzmann: 
The Man Who Trusted Atoms", Oxford Uni. Press, 1998.} 



\bibitem[{\citenamefont{Boltzmann}(1968)}]{boltzmann68}
  \bibinfo{author}{\bibfnamefont{Boltzmann}, \bibnamefont{Ludwig}},
  \bibinfo{year}{1968}, {``}\bibinfo{title}{Studien uber das gleichgewicht der lebendigen kraft zwischen bewegten ma-teriellen},{''} \bibinfo{journal}{Wien. Ber.} \textbf{\bibinfo{volume}{58}},  \bibinfo{pages}{517}.




\bibitem[{\citenamefont{Gallavotti}(1994)}]{boltz84}
\bibinfo{author}{\bibnamefont{Gallavotti}, \bibfnamefont{G.}},
  \bibinfo{year}{1995}, {``}\bibinfo{title}{Ergodicity, ensembles, irreversibility in Boltzmann and beyond},{''} \bibinfo{journal}{J. Stat. Phys.}
  \textbf{\bibinfo{volume}{78}},  \bibinfo{pages}{1571}.




\bibitem{bri} Birkhoff G. D., 1931,  ``Proof of the ergodic theorem", Proc. Natl. Acad. Sci USA, {\bf 17} (12), 656.



\bibitem{khinchin} Khinchin, A. I., 1960, {\it "Mathematical foundations of statistical mechanics"}, Dover.


\bibitem{livi} Livi R., Pettini M., Ruffo S., and Vulpiani A., 1987, J. Stat. Phys. {\it 48}, 539. 


\bibitem{bookvul} Castiglione P., Falcioni M., Lesne A., and Vulpiani A., 2008,  {\it ``Chaos and coarse graining in statistical mechanics"}, Cambridge University Press, UK.



\bibitem{jaynes79} Jaynes, E. T., 1979,``Where do we stand on maximum entropy", in ``Maximum entropy formalism, R.D. Levine and M. Tribus (editors)", MIT press, Cambridge, MA, p15.


\bibitem{rigolnature} Rigol M., Dunjko V., Olshani M., 2008, ``Thermalization and its mechanism for generic isolated quantum system" , Nature, {\bf 452}, 854.




\bibitem{rigol2009a} M. Rigol, "Breakdown of thermalization in finite one-dimensional systems", Phys. Rev. Lett. {\bf 103}, 100403 (2009).



\bibitem{rigol2009b} M. Rigol, "Quantum quenches and thermalization in one-dimensional fermionic systems", Phys. Rev. A {\bf 80}, 053607 (2009).


\bibitem{deutsch} Deutsch J. M., 1991, ``Quantum statistical mechanics in a closed system", Phys. Rev. A, {\bf 43}, 2046.



\bibitem{sred94} Srednicki M., 1994, ``Chaos and quantum thermalization", Phys. Rev. E, {\bf 50}, 888; see also arXiv:cond-mat/9406056, and arXiv:cond-mat/9410046.




\bibitem{gltz} Goldstein S., Lebowitz J., Tumulka R., Zanghi N., ``Long-time behaviour of macroscopic quantum systems: commentary accompanying the English translation of John von Neumann's 1929 article on the quantum ergodic theorem"
 

\bibitem{neumann} John von Neumann, "Proof of the Ergodic Theorem and the H-Theorem in Quantum Mechanics", Eur. Phys. J. H {\bf 35}: 201-237 (2010), arXiv:1003.2133, [English translation by Roderich Tumulka of "Beweis des Ergodensatzes und des H-Theorems" in  Zeitschrift fuer Physik {\bf 57}: 30-70 (1929)].



\bibitem{normal} Sheldon Goldstein, Joel L. Lebowitz, Christian Mastrodonato, Roderich Tumulka, Nino Zanghi, "Normal Typicality and von Neumann's Quantum Ergodic Theorem", Proceedings of the Royal Society A {\bf 466}, 3203 (2010). 


\bibitem{sheldon2006} Sheldon Goldstein, Joel L. Lebowitz, Roderich Tumulka, Nino Zanghi, "Canonical Typicality", Phys. Rev. Lett. {\bf 96}, 050403 (2006). 



\bibitem{s52} E. Schr\"{o}dinger, {\it "Statistical thermodynamics"}, 2nd Edition, CUP, 1952.



\bibitem{mahler} Gemmer  J., Mahler  G., "Distribution of Local Entropy in the Hilbert Space of Bi-partite Quantum Systems: Origin of Jaynes’ Principle", Euro. Phys. J. B {\bf 31}, 249-257 (2003). quant-ph/0201136.




\bibitem{tasaki}Tasaki, H., "From Quantum Dynamics to the Canonical Distribution: General Picture and a Rigorous Example",  Phys. Rev. Lett. {\bf 80}, 1373-1376 (1998), cond-mat/9707253.


\bibitem{lindenpre79} N. Linden, S. Popescu, A. J. Short, and A. Winter, "Quantum mechanical evolution towards thermal equilibrium", Phys. Rev. E {\bf 79}, 061103 (2009).


\bibitem{cho} J. Cho and M. S. Kim, "Emergence of canonical ensembles from pure quantum states", Phys. Rev. Lett. {\bf 104}, 170402 (2010).


\bibitem{sandu} S. Popescu, A. J. Short, A. Winter, "Entanglement and the foundation of statistical mechanics",  Nature Physics {\bf 21}, 754–758 (2006); arXiv:quant-ph/0511225.




\bibitem{nexp} R. V. Jensen and R. Shanker, "Statistical Behavior in Deterministic Quantum Systems with Few Degrees of Freedom", Phys. Rev. Lett., {\bf 54}, 1879 (1985); K. Saito, S. Takesue, and S. Miyashita, "System-Size Dependence of Statistical Behavior in Quantum System",  J. Phys. Soc. Jpn. {\bf 65}, 1243 (1996); "Thermal conduction in a quantum system", Phys. Rev. E {\bf 54}, 2404 (1996).


\bibitem{reimann} Peter Reimann, "Typicality for generalized microcanonical ensembles", Phys. Rev. Lett. {bf 99}, 160404 (2007).


\bibitem{dntreimann} A version of DNT (equilibration as opposed to thermalization)  has been proved in\cite{reimann2008}.  Here it is shown that the time average of the difference in the instantaneous expectation values of an observable and its ensemble average is very small. Ensemble here is the diagonal ensemble (equation 11). In literature it is called equilibration\cite{lindenpre79,reimann2008}. The agreement of instantaneous expectation value of an observable with microcanonical or canonical ensemble is called thermalization which is a much more difficult problem and possibly involves chaos.


\bibitem{reimann2008} Peter Reimann, "Foundations of statistical mechanics under experimently realistic conditions", Phys. Rev. Lett. {bf 101}, 190403 (2008).


\bibitem{hal}Hal Tasaki, "The approach to thermal equilibrium and "thermodynamic normality"---An observation based on the works by Goldstein, Lebowitz, Mastrodonato, Tumulka, and Zanghi in 2009, and by von Neumann in 1929",
 arXiv:1003.5424.



\bibitem{peres} Asher Peres, "Ergodicity and mixing in quantum theory. I", Phys. Rev. A. {\bf 30}, 504 (1984).


\bibitem{gut} M. C. Gutzwiller, "Periodic Orbits and Classical Quantization Conditions", J. Math. Phys. {\bf 12}, 343 (1971).


\bibitem{balian}R. Balian and C. Bloch, "Distribution of eigenfrequencies for the wave equation in a finite domain: III. Eigenfrequency density oscillations",Ann. Phys.(NY) {\bf 69}, 76 (1972); "Solution of the Schrödinger equation in terms of classical paths", {bf 85}, 514 (1974).


\bibitem{berry}M. V. Berry and M. Tabor, "Closed Orbits and the Regular Bound Spectrum",  Proc. R. Soc. London, Ser. A {\bf 349} 101 (1976).



\bibitem{fermi} Enrico Fermi, "{\it Themodynamics}" Dover Publications, 1956.


\end{thebibliography}
\end{document}